\documentclass[conference,a4paper]{IEEEtran}
\IEEEoverridecommandlockouts
\usepackage[utf8]{inputenc}
\usepackage{xcolor}
\usepackage{balance}
\usepackage{bm}
\usepackage{cite}
\usepackage{amsmath,amssymb,amsfonts,bm}
\usepackage{graphicx}
\usepackage{textcomp}
\usepackage{xcolor}
\usepackage{comment}
\usepackage{amsmath,amssymb,amsfonts}
\usepackage{tikz}
\usetikzlibrary{arrows,automata,positioning}
\usetikzlibrary{arrows.meta}
\usepackage{graphicx}
\usepackage{textcomp}
\usepackage{xcolor}
\usepackage{lipsum}
\usepackage{booktabs}
\usepackage{color, colortbl}
\usepackage{algorithm}
\usepackage{algpseudocode}
\usepackage{dsfont}
\usepackage[acronyms]{glossaries}
\usepackage{tcolorbox}
\usetikzlibrary{calc, positioning, arrows.meta, fit}

\pgfdeclarelayer{lay00}
\pgfdeclarelayer{back00}
\pgfdeclarelayer{lay01}
\pgfdeclarelayer{back01}
\pgfdeclarelayer{lay02}
\pgfdeclarelayer{back02}
\pgfdeclarelayer{lay03}
\pgfdeclarelayer{back03}
\pgfdeclarelayer{back04}
\pgfdeclarelayer{lay10}
\pgfdeclarelayer{back10}
\pgfdeclarelayer{lay11}
\pgfdeclarelayer{back11}
\pgfdeclarelayer{lay12}
\pgfdeclarelayer{back12}
\pgfdeclarelayer{lay13}
\pgfdeclarelayer{back13}
\pgfdeclarelayer{back14}
\pgfdeclarelayer{lay30}
\pgfdeclarelayer{back30}
\pgfdeclarelayer{lay31}
\pgfdeclarelayer{back31}
\pgfdeclarelayer{lay32}
\pgfdeclarelayer{back32}
\pgfdeclarelayer{lay33}
\pgfdeclarelayer{back33}
\pgfdeclarelayer{back34}

\pgfsetlayers{back34,back33,lay33,back32,lay32,back31,lay31,back30,lay30,back14,back13,lay13,back12,lay12,back11,lay11,back10,lay10,back04,back03,lay03,back02,lay02,back01,lay01,back00,lay00,main}

\newlength{\figSpaceX}
\newlength{\figSpaceY}
\newlength{\nodeSize}
\colorlet{group color}{white}
\colorlet{components color}{black!15!white}

\ifCLASSOPTIONcompsoc
 \usepackage[caption=false,font=normalsize,labelfont=sf,textfont=sf]{subfig}
\else
 \usepackage[caption=false,font=footnotesize]{subfig}
\fi
\begin{document}
%
\title{
 Variational Bayesian Estimation of Low Earth Orbits for Satellite Communication
}

\author{\IEEEauthorblockN{
Anders Malthe Westerkam\IEEEauthorrefmark{1}\IEEEauthorrefmark{4},   
Amélia Struyf\IEEEauthorrefmark{2}\IEEEauthorrefmark{4},   
Dimitri Lederer\IEEEauthorrefmark{3}, 
Troels Pedersen\IEEEauthorrefmark{1}, 
François Quitin\IEEEauthorrefmark{2} 
}                                     

\IEEEauthorblockA{\IEEEauthorrefmark{1}
Dept. of Electronic systems AAU, Aalborg, Denmark \{amw,troels\}@es.aau.dk}
\IEEEauthorblockA{\IEEEauthorrefmark{2}
Brussels School of Engineering, Univeristé Libre de Bruxelles, Brussels, Belgium, \{amelia.struyf, francois.quitin\}@ulb.be}
\IEEEauthorblockA{\IEEEauthorrefmark{3}
ICTEAM, UCLouvain, Louvain-la-Neuve, Belgium, 
dimitri.lederer@uclouvain.be}
\thanks{\IEEEauthorrefmark{4} Amélia Struyf and Anders Malthe Westerkam are co-first authors.}
}



\maketitle

\begin{abstract}
Low-earth-orbit (LEO) satellite communication systems that use millimeter-wave frequencies rely on large antenna arrays with hybrid analog-digital architectures for rapid beam steering. LEO satellites are only visible from the ground for short periods of times (a few tens of minutes) due to their high orbital speeds. This paper presents a variational message passing algorithm for joint localization and beam tracking of a LEO satellite from a ground station equipped with a hybrid transceiver architecture. The algorithm relies on estimating the parameters of the orbit, which is modelled as circular. Angles are then obtained from the orbit in a straightforward manner. Simulation results show that the proposed method is highly resilient to missed detections, enables reliable satellite tracking even near the horizon, and effectively alleviates the ambiguities inherent in hybrid architectures.  

\end{abstract}

\vskip0.5\baselineskip

%

\section{Introduction}
Satellite based communication systems hinges on the ability to accurately steer the beam of the ground stations (GS) antenna to the direction of the satellite. In general antennas with high directivity are employed in a quest to minimize transmission losses and at the same time avoid interference. Considering non-geostationary satellites, the ground station should be able to track the satellite movement close enough to avoid losing connection to the satellite.   This is a particularly pressing concern in communication systems employing Low Earth Orbit (LEO) satellites. In such cases, the low orbit, in combination with high speed, result in rapidly changing direction and very brief visibility windows, typically on the order of ten minutes during each orbital pass.  Hence, effective beam management is crucial to ensure performance of LEO communication systems.
For LEO communication, mechanically steered high-gain parabolic antennas are depreciated due to their slow steering and unflexible antenna beams. In contrast, electronically steered antenna arrays offer much greater speed and flexibility.  Such arrays enable fast steering of multiple beams toward multiple satellites as well as null steering, both of which are important features for LEO satellite constellations \cite{sturdivat}.  Typically, hybrid analog–digital architectures are employed to reduce cost compared to fully digital arrays while preserving significant beamforming capabilities \cite{HB_ref, zhang}.
Research on beam management for LEO systems has mainly focused on the satellite side, while less attention has been paid to user equipment and GS.  Existing beam tracking solutions for millimeter wave (mmWave) systems typically rely on continuous beam tracking \cite{Artiga} or two-step methods combining angle-of-arrival (AoA) estimation with Kalman filtering \cite{Wang}. Both suffer from limited robustness in the case of signal blockage or interruption. Moreover, traditional AoA estimation techniques applied to hybrid architectures suffer from phase ambiguity when the phase center the sub-arrays of the transceiver is greater than $\lambda/2$ \cite{Huang}. 

\begin{figure}
    \centering
\includegraphics[width=0.58\linewidth]{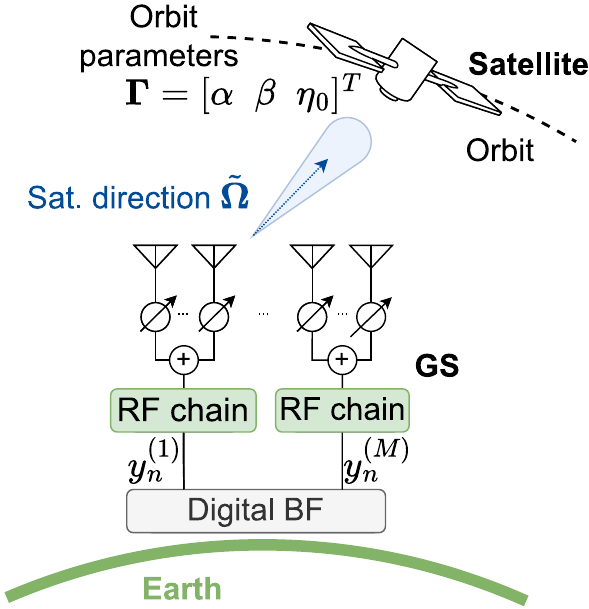}
\caption{Considered scenario with a GS tracking a LEO satellite.}
\label{fig:system_model}
\end{figure}
It is well-known that the information passed from one step to the next should be a sufficient statistic--- a condition which is generally hard to ensure and is not met in traditional two-step tracking schemes such as direction estimation followed by Kalman filtering. This motivates the research interest in joint direction estimation and tracking in related applications such as drone tracking in radar systems \cite{Westerkam2025,Kitchen2025}. In these works Bayesian Variational message passing (VMP)  is employed to develop joint estimation and tracking methods which  has been shown to improve robustness and accuracy compared to traditional multi-step methods. 

The present contribution develops a novel VMP algorithm for joint localization and beam tracking of a LEO satellite from a GS with a hybrid architecture at mmWave frequencies. The proposed method estimates the parameters of the satellite orbit, which is modelled as circular, and computes the AoA based on these estimates. We assume a scenario in which the GS has no access to prior information about the satellite position. Simulation results highlight the advantages of this joint approach compared to conventional methods.

\section{System and orbit model}

\subsection{System model}
Consider the downlink transmission between a LEO satellite and a GS illustrated in Fig.~\ref{fig:system_model}. As shown in Fig.~\ref{fig:Orbitfig}, the satellite orbit is specified with respect to a  a right-handed coordinate system at the GS, such that the z-axis points to zenith. At time $t$, the position of the satellite is given by the 3D vector $\bm{\Omega}(t;\bm{\Gamma})$ specified by the orbit parameter vector  $\bm{\Gamma}$. Thus the unit vector $\tilde{\bm{\Omega}}(t)=\bm{\Omega}(t;\bm{\Gamma})/||\bm{\Omega}(t;\bm{\Gamma})||$ is the  the satellite direction at time $t$. We assume that the direction varies slow enough to be considered constant over a recording time interval $n\Delta t \leq t < (n+1)\Delta t$, i.e., $\tilde{\bm{\Omega}}(t) =\tilde{\bm{\Omega}}_n.$ 

The GS is equipped with a Uniform Planar Array (UPA) with a hybrid analog-digital architecture with infinite resolution fixed-amplitude phase shifters and $ \lambda/2$-spaced antennas. We consider a partially-connected hybrid architecture, with $M$ RF chains each connected to a different subset of $N_{sub}$ antennas. Because of this independent sub-array architecture, the signals after analog combining do not meet the Nyquist spatial sampling criterion, leading to a $2 \pi$ phase ambiguity \cite{Huang}. The satellite is equipped with a UPA with an analog architecture with $N_t$ $\lambda /2$-spaced antennas, with its beam assumed to always perfectly point towards the GS with maximum gain. 

The signal received by the $m^{th}$ subarray at time instant $n$ is modeled as
\begin{equation}
   \mathbf{y}^{(m)}(\bm{\Omega}_n)\hspace{-2.5pt} =\hspace{-2.5pt} (\bm{b}^{(m)}(\tilde{\bm{\Omega}}_n))^H \bm{a}^{(m)}(\tilde{\bm{\Omega}}_n) f^{(m)}(\tilde{\bm{\Omega}}_n) h(\bm{\Omega}_n) \bm{s}+ \bm{v}_{n}^{(m)},
\end{equation}
where $\bm{b}^{(m)}(\tilde{\bm{\Omega}}_n) \in \mathbb{C}^{N_{sub} \times 1}$  corresponds to the analog combining weights, $\bm{a}^{(m)}(\tilde{\bm{\Omega}}_n) \in \mathbb{C}^{N_{sub} \times 1}$  is the receive steering vector, $f^{(m)}(\tilde{\bm{\Omega}}_n) \in \mathbb{C}^{1 \times 1}$ is the phase shift between the first and the $m^{th}$ subarray,  $h(\bm{\Omega}_n)  \in \mathbb{C}^{1 \times 1} $ represents the channel gain, $\bm{s} \in \mathbb{C}^{N_{s} \times 1}$ corresponds to the transmitted signal and $\bm{v}_{m}  \in \mathbb{C}^{N_{s} \times 1}$ is complex circular symetric distributed noise $\bm{v}_m\sim N^C(0,\bm{\Lambda}_v)$, where $N^C(\bm{\mu},\bm{\Lambda})$ denotes the complex circular symmetric distribution with mean $\bm{\mu}$ and precision matrix $\bm{\Lambda}$. For brevity, the explicit dependence on the AoA $\bm{\Omega}_n$ is dropped, and the time dependence is noted by a subscript, for example, $h(\bm{\Omega}_n) = h_n$. By stacking received signals of all $M$ subarrays in a single vector, we obtain
\begin{equation}
\mathbf{y}_{n} = h_{n}\,\mathbf{x}_{n} + \bm{v}_{n}.
\label{eq:compact}
\end{equation}
 Here, $\mathbf{x}_n$ corresponds to received signals that include beamforming effects arising from a satellite in direction $\tilde{\bm{\Omega}}_n$ without the channel effects.
An Orthogonal Frequency Division Multiplexing (OFDM) system is considered, where each subcarrier is represented by a narrowband line-of-sight (LOS) channel. Multipath effects are neglected in this scenario and the Doppler shift due to satellite motion is assumed to be perfectly compensated for at the receiver. Under these assumptions, the channel model is $h_n =  \rho_n e^{j \chi_{n}}$
where $\chi_n\sim U(0,2\pi)$ and $\rho_n$ includes the attenuation due to the free space path loss, the atmospheric attenuation, and the gains of the transmit and receive antenna arrays. In this work, atmospheric attenuation is set according to models provided by the ITU recommendation~\cite{ITU-RP61813}. 

\subsection{Orbit Model}
\begin{figure}
    \centering
    \includegraphics[width=0.7\linewidth]{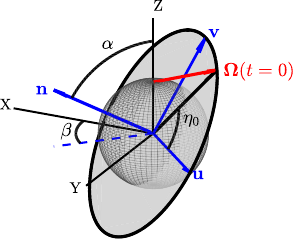}
    \caption{Satellite orbit parameters $\bm{\Gamma}$ and satellite direction $\bm{\Omega}$. }
    \label{fig:Orbitfig}
\end{figure}

Satellite orbits can be complicated and are affected by factors such as atmospheric drag, non-uniform Earth gravity, and the gravitational pull of other celestial bodies \cite{orbit}. However, considering a single revolution, we approximate the orbit as circular. Considering the orbit to have fixed and known angular velocity, $\omega$ and radius $R$. The orbit can then be described by the vectors $\bm{u}$ and $\bm{v}$ as shown in Fig.~\ref{fig:Orbitfig}. These vectors can be defined in terms of two rotation angles $\bm{u} = R[-\sin(\beta),\, \cos(\beta),\, 0]^{T}$ and 
$\bm{v} = R[-\cos(\alpha)\cos(\beta),\, -\cos(\alpha)\sin(\beta),\, \sin(\alpha)].^{T}$  

The time evolution of the satellite position (as seen from the GS) $\bm{\Omega}$ is then given by
\begin{equation}\label{eq:AoAcalc}
    \bm{\Omega}(t;\bm{\Gamma}) = \bm{u}\cos(\omega t-\eta_0) + \bm{v}\sin(\omega t-\eta_0) - \begin{bmatrix}
        0&0& h_{e}
    \end{bmatrix}^T
\end{equation}
where $\eta_0$ is the starting angle and $h_{e}$ is the height of the orbit as seen from the Earth.
As such, the vector of orbit parameters to be estimated is $\bm{\Gamma} = \begin{bmatrix}\alpha& \beta& \eta_0\end{bmatrix}^T$.

\section{Probabilistic model}
\subsection{Probabilistic model}
We assume that the orbit parameters $\bm{\Gamma}$ are constant but unknown, and the channel $h_n$ is unknown but time varying. We define $\mathbf{y}_{0:N} = \begin{bmatrix}\mathbf{y}_0 & \mathbf{y}_1 & \hdots & \mathbf{y}_
N\end{bmatrix}$, and  $\bm{h}_{0:N} = \begin{bmatrix}h_0 & h_1 & \hdots & h_
N\end{bmatrix}$; by assuming the signal vectors are conditionally independent given $\bm{\Gamma}$ and $\bm{h}_{0:N}$, the posterior probability up to a constant is given by
\begin{equation}\label{eq:posterior}
    p(\bm{h}_{0:N},\bm{\Gamma}|\mathbf{y}_{0:N}) \propto p(\bm{\Gamma})\prod_{n=0}^N p(h_n) p(\mathbf{y}_n|h_n,\bm{\Gamma}).
\end{equation}
Conditioned on $\bm{\Gamma}$, $\mathbf{x}_n$ is fully defined and hence the distribution $p(\mathbf{y}_n|h_n,\bm{\Gamma})$ is a Gaussian
\begin{equation}
    p(\mathbf{y}_n|\bm{\Gamma}) = N^C(h_n\mathbf{x}(\tilde{\bm{\Omega}}_n(\bm{\Gamma})),\bm{\Lambda}_v).
\end{equation}
For brevity $\mathbf{x}(\tilde{\bm{\Omega}}_n(\bm\Gamma))$ will be shorted to $\mathbf{x}_n(\bm{\Gamma})$. The channel prior is  a zero mean circular symmetric Gaussian, i.e.,
\begin{equation}
    p(h_n) = N^C(0,\gamma_p).
\end{equation}
The prior on the orbit parameters  will be considered uniform; however, not all combinations of $\alpha,\beta,\eta_0$ yield an orbit that passes above the horizon, hence we limit the distribution to such orbits while otherwise treating each parameter as independent,
\begin{equation}\label{eq:prior_Gamma}
    p(\bm{\Gamma})=\begin{bmatrix}
            U(1.25,1.87)&\phantom{m} U(0,2\pi)  & U(0,2\pi) \end{bmatrix}^T.
\end{equation}

\subsection{Variational Message Passing}
We approximate the intractable  posterior in \eqref{eq:posterior} by the mean field approach, cf \cite{bishop2007}. The posterior surrogate  
\begin{equation}\label{eq:full_sorogate}
    q(\bm{\Gamma},h_{0:N})  = q(\bm{\Gamma})\prod_{n=0}^{N}q(h_n),
\end{equation}
is  chosen to  minimize the Kullback-Leibler divergence with respect to the true posterior \eqref{eq:posterior}. Following \cite{bishop2007}, this is achieved with surrogates  
\begin{equation}\label{eq:qh}
    \ln q(h_n) = \mathbb{E}_{\bm{\Gamma}}[\ln p(\mathbf{y}_n|h_n,\bm{\Gamma})] + \ln p(h_n) + \text{const},
\end{equation}
\vspace{-15pt}
\begin{equation} \label{eq:qGamma}
    \ln q(\bm{\Gamma}) = \sum_{n=0}^{N}\mathbb{E}_{\backslash \bm{\Gamma}}[\ln p(\mathbf{y}_n|h_n,\bm{\Gamma})] + \ln p(\bm{\Gamma}) + \text{const}.
\end{equation}

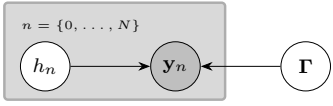
\begin{figure}
    \centering
    \resizebox{0.5\columnwidth}{!}{%
    \begin{tikzpicture}[variable node/.style={circle,yshift=0.2,draw,minimum width=\nodeSize,inner sep=0,font=\footnotesize,fill=white},
	deterministic node/.style={circle,dashed,draw,minimum width=\nodeSize,font=\footnotesize,fill=white}, data node/.style={circle,yshift=0.2,draw,minimum width=\nodeSize,inner sep=0,font=\footnotesize,fill=lightgray},
	pdf/.style={draw,fill=white,font=\scriptsize,rounded corners=2pt},
	scale=1,>={Stealth[scale=1]}
	]

\node[variable node] at (2.3\figSpaceX,0\figSpaceY) (Gamma){$\boldsymbol{\Gamma}$};
\node[data node] at (1.5\figSpaceX,0\figSpaceY) (yn) {$\mathbf{y}_n$};

\node[variable node] at (0.7\figSpaceX,0\figSpaceY) (hn){$h_n$};

\node[anchor=base west] at (0.5\figSpaceX,0.6\figSpaceY) (ndenote) {\tiny $n=\{0,\hdots,N\}$};

\node at (1.7\figSpaceX,0\figSpaceY) (dummynode){};

\draw[->] (Gamma)--(yn);
\draw[->] (hn)--(yn);

\begin{pgfonlayer}{back01}
                \node[fill=components color,draw=components color!80!black, rounded corners=0.5mm, line width=1pt,fit=(yn) (hn) (ndenote) (dummynode)] (0pt,0pt) {};                
\end{pgfonlayer}
\end{tikzpicture}
}
    \caption{%
   Bayesian network for Low Earth Orbit estimation. Each term in Eq. (\ref{eq:qGamma}) and Eq. (\ref{eq:qh}) can be seen as a message passed along the edge of the graph.
      }
    \label{fig:Baysian_graph}
\end{figure}
Here $\mathbb{E}_{\backslash(\cdot)}$ denotes the expectation with respect to all parameters except $(\cdot)$.
Each term can be seen as a message passed along the edge of the graph in Fig.~\ref{fig:Baysian_graph}. The first term in \eqref{eq:qh} is,
\begin{multline}\label{eq:exp_over_gamma}
    \mathbb{E}_{\bm{\Gamma}}[\ln p(\mathbf{y}_n|h_n,\bm{\Gamma})] = -\langle h_n|\mathbb{E}_{\bm{\Gamma}}[\langle\mathbf{x}_n(\bm{\Gamma})|\bm{\Lambda}_v|\mathbf{x}_n (\bm{\Gamma})\rangle]|h_n\rangle \\+ 2Re\{\langle\mathbf{y}_n|\bm{\Lambda}_v|\mathbb{E}[\mathbf{x}_n(\bm{\Gamma})\rangle|h_n\rangle]\} + \text{const},
\end{multline}
where $\langle\mathbf{a}|\mathbf{M}|\mathbf{b}\rangle=\mathbf{a}^H\mathbf{M}\mathbf{b}$ is the bra-ket notation. Eq. \eqref{eq:exp_over_gamma} is the functional form of a complex circular symmetric Gaussian. Combining this with the prior $p(h_n)$, the surrogate distribution is also complex Gaussian, 
    $q(h_n) = N^C(\bar{h}_n,\bar{\bar{h}}_n),$
where $\bar{\cdot}$ denotes the mean, and $\bar{\bar{\cdot}}$ denotes the (co)variance, given by,
\begin{equation}\label{eq:mean_of_h}
    \bar{h}_n = \bar{\bar{h}}_n\langle\mathbb{E}_{\bm{\Gamma}}[\mathbf{x}_n(\bm{\Gamma})]|\bm{\Lambda}_v|\mathbf{y}_n\rangle,
\end{equation}
\begin{equation}\label{eq:variance_of_h}
    \bar{\bar{h}}_n = (\mathbb{E}_{\bm{\Gamma}}[\langle\mathbf{x}_n(\bm{\Gamma})|\bm{\Lambda}_v|\mathbf{x}_n (\bm{\Gamma})\rangle]+\gamma_p)^{-1}.
\end{equation}
Employing the delta method, the expectations can be approximated as 
\begin{equation}
    \mathbb{E}_{\bm{\Gamma}}[\mathbf{x}_n(\bm{\Gamma})] \approx \mathbf{x}_n(\bar{\bm{\Gamma}}),
\end{equation}
\begin{multline}
    \mathbb{E}_{\bm{\Gamma}}[\langle\mathbf{x}_n(\bm{\Gamma})|\bm{\Lambda}_v|\mathbf{x}_n (\bm{\Gamma})\rangle] \approx \langle\mathbf{x}_n(\bar{\bm{\Gamma}})|\bm{\Lambda}_v|\mathbf{x}_n(\bar{\bm{\Gamma}})\rangle \\+ Tr(\bar{\bar{\bm{\Gamma}}} \langle\nabla_{\bm{\Gamma}}\mathbf{x}_n(\bm{\Gamma})\bigg|_{\bar{\bm{\Gamma}}}|\bm{\Lambda}_v|\nabla_{\bm{\Gamma}}\mathbf{x}_n(\bm{\Gamma})\bigg|_{\bar{\bm{\Gamma}}}\rangle),
\end{multline}
where $\nabla_{\bm{\Gamma}}$ denotes the gradient with respect to  $\bm{\Gamma}$.

For the surrogate $q(\bm{\Gamma})$ in \eqref{eq:qGamma}, the first term reads 
\begin{multline}\label{eq:exp_for_qgamma}
    \mathbb{E}_{\backslash \bm{\Gamma}}[\ln p(\mathbf{y}_n|h_n,\bm{\Gamma})] = -(|\bar{h}_n|^2+\bar{\bar{h}}_n)\langle \mathbf{x}_n(\bm{\Gamma})|\bm{\Lambda}_v|\mathbf{x}_n(\bm{\Gamma})\rangle \\+ 2Re\{\langle \mathbf{y}_n|\bm{\Lambda}_v|\bar{h}_n\mathbf{x}_n(\bm{\Gamma})\rangle\} + \text{const.}
\end{multline}
The surrogate is obtained by inserting \eqref{eq:exp_for_qgamma} into \eqref{eq:qGamma}. The resulting surrogate is highly sensitive to the estimation accuracy of the phase of the channel, due to the real part in the second term. To remedy this, we replace the real operator by an absolute operator as proposed in \cite{Kitchen2025}. Accordingly, \eqref{eq:exp_for_qgamma} is replaced by 
\begin{multline}\label{eq:exp_for_qgamma_approx}
    \mathbb{E}_{\backslash \bm{\Gamma}}[\ln p(\mathbf{y}_n|h_n,\bm{\Gamma})] = -(|\bar{h}_n|^2+\bar{\bar{h}}_n)\langle \mathbf{x}_n(\bm{\Gamma})|\bm{\Lambda}_v|\mathbf{x}_n(\bm{\Gamma})\rangle \\+ |\langle \mathbf{y}_n|\bm{\Lambda}_v|\bar{h}_n\mathbf{x}_n(\bm{\Gamma})\rangle| + \text{const.}
\end{multline}
This does not yield a known distribution in $\bm{\Gamma}$, as such, we choose to approximate the surrogate $q(\bm{\Gamma})$ by a multivariate Gaussian with mean, $\bar{\bm{\Gamma}}$, approximated by the mode of $q(\bm{\Gamma})$ which can be found numerically. The covariance, $\bar{\bar{\bm{\Gamma}}}$, is approximated using the Fisher information matrix evaluated at the mean $\mathcal{I}(\bar{\bm{\Gamma}})$ as this is the lower bound on the variance of an estimator $\Sigma_\theta\geq 1/\mathcal{I}(\theta)$. Hence, the moments are
\begin{equation}\label{eq:mean_of_gamma}
    \bar{\bm{\Gamma}} = \underset{\bm{\Gamma}}{\arg\max}\, \ln q(\bm{\Gamma}),\phantom{m} \bar{\bar{\bm{\Gamma}}} = \bigg[-\nabla\nabla^T\ln q(\bm{\Gamma})\bigg|_{\bar{\bm{\Gamma}}}\bigg]^{-1}.
\end{equation}
Here, $\nabla\nabla^T$ denotes the Hessian, which is found numerically.
\subsection{Algorithm}
To initialize the numeric optimization for the estimate of $\bar{\bm{\Gamma}}$ we utilize the posterior on the orbit parameters, $p(\bm{\Gamma}|\bm{\Omega}_0)$ which can significantly limit the space for optimization. As the functional form is intractable we sample from the distribution instead. Here we use an approximate Bayesian computation (ABC) algorithm as outlined in Alg.~\ref{alg:samp} to generate the set of samples here denoted $\bm{\tilde{\bm{\Gamma}}}$. It is expected that some of these samples fall within the main lobe; however, with just one sample, the problem is still fully ambiguous, as an infinite number of orbits will pass through a single point. However not all of these orbits are of interest to track and we limit the considered set by requiring the orbits to stay above the horizon in time step two, i.e., ${\Omega}_z(\Delta t;\bm{\Gamma}_l)>0$ where ${\Omega}_z$ denotes the projection unto the z-axis. Furthermore, we consider only orbits which are rising, $v_\theta<0$, the velocity of $\tilde{\bm{\Gamma}}$ is estimated from \eqref{eq:AoAcalc} using finite difference. These two conditions are implemented in Alg.~\ref{alg:samp}. The points, $\tilde{\bm{\Gamma}}_{1:N_{acc}}$, are used to approximate the prior $p(\bm{\Gamma})$ in \eqref{eq:qGamma} by kernel density estimator using gaussian kernels with bandwidth 0.005. The initial mean estimate, $\bar{\bm{\Gamma}}_1$, is then found by optimizing $\ln q({\bm{\Gamma}})$ starting in the 60 most likely points from $\tilde{\bm{\Gamma}}_{1:N_{acc}}$ and choosing the best. Subsequent optimization steps are initialized at the current $\bar{\bm{\Gamma}}$

One parameter of special interest is the time between updates of the surrogate $q(\bm{\Gamma})$, as the information gained about the orbit is tied to observing the signal arriving at different AoAs. We found that using time updates of 20 seconds yield satisfying results. The AoA estimation for beam steering is inferred from the current $q(\bm{\Gamma})$
using the transform \eqref{eq:AoAcalc}. It is important to mention that the AoA can be inferred in continuous time such that the communication with the satellite can occur more frequently.
Additionally the proposed algorithm removes $2\pi$ phase ambiguity associated with hybrid architectures by combining multiple signals. The full algorithm is outlined in Alg.~\ref{alg:cap}.
\begin{algorithm}\small
\caption{ABC algorithm for generating $\bm{\Gamma}\sim p(\bm{\Gamma}|\hat{\bm{\Omega}}_0)$}\label{alg:samp}
\begin{algorithmic}
\State \textbf{Input:} Initial AoA estimate $\hat{\tilde{\bm{\Omega}}}_0$ and desired number of samples $N_{samp}$
\State \textbf{Output:} $N_{samp}$ gamma samples $\tilde{\bm{\Gamma}}_{1:N_{samp}}$

\State $N_{trials} \leftarrow 10^6$, $l\leftarrow1$  

\While{$l <N_{trials}$ } 
    \State $ \alpha_l \sim U(1.25, 1.87),\,  \beta_l\sim U(0,2\pi),\,$ $\eta_{0,l}\sim U(0,2\pi)$
    \If{$\bm{\Omega}_z(\Delta t;\bm{\Gamma}_l)>0$ and $v_\theta<0$}
    \State $\bm{\Gamma}_l \leftarrow\begin{bmatrix}
        \alpha_l&\beta_l&\eta_{0,l}        
    \end{bmatrix}^T$
    \State $\tilde{\bm{\Omega}}_l \leftarrow \bm{\Omega}_0(\bm{\Gamma}_l)/||\bm{\Omega}_0(\bm{\Gamma}_l)||$, $f_l \leftarrow \tilde{\bm{\Omega}}_l^T\hat{\tilde{\bm{\Omega}}}_0$, $l\leftarrow l+1$    
    \EndIf    
        
\EndWhile
\State Let $\tilde{\bm{\Gamma}}_{1:N_{samp}}$ be the $N_{samp}$ Gammas with the highest $f_l$
\end{algorithmic}
\end{algorithm}
\vspace{-20pt}
\begin{algorithm}
\small
\caption{Orbit estimation algorithm}\label{alg:cap}
\begin{algorithmic}
\State \textbf{Input:} Signal vectors $\mathbf{y}_{0:N}$, and initial AoA estimate $\hat{\tilde{\bm{\Omega}}}_0$
\State \textbf{Output:} The first and second moments of the channel  $\bar{h}_{1:N},\bar{\bar{h}}_{1:N}$ and the orbit parameters $\bar{\bm{\Gamma}}_{1:N},\bar{\bar{\bm{\Gamma}}}_{1:N}$
\State \textbf{Initialization}:
\State Generate $\tilde{\bm{\Gamma}}_{1:N_{accept}}$ using Alg.~\ref{alg:samp}, update $q(\bm{\Gamma})$ using $\mathbf{y}_{0:1}$
\State $\bar{\bm{\Gamma}}_1 \leftarrow \arg\max_{\bm{\Gamma}} \ln q(\bm{\Gamma})$ starting in $\tilde{\bm{\Gamma}}_{1:N_{acc}}$
\State $\bar{\bar{\bm{\Gamma}}}_1^{-1} \leftarrow -\nabla\nabla^T\ln q(\bm{\Gamma})\bigg|_{\bar{\bm{\Gamma}}_1}$
\State \textbf{Main algorithm:}
\For{$n\leftarrow2:N$}
\State Calculate $\{\bar{\bar{h}}_n$,$\bar{h}_n\}$ using $\{\bar{\bm{\Gamma}}_{n-1},$ $\bar{\bar{\bm{\Gamma}}}_{n-1}\}$ in \eqref{eq:variance_of_h} and \eqref{eq:mean_of_h}
\State Update $\ln q(\bm{\Gamma})$ using $\mathbf{y}_n$, $\bar{\bar{h}}_n$ and $\bar{h}_n$
\State $\bar{\bm{\Gamma}}_n\leftarrow$argmax$_{\bm{\Gamma}}$ $\ln q(\bm{\Gamma})$ starting in $\bar{\bm{\Gamma}}_{n-1}$
\State $\bar{\bar{\bm{\Gamma}}}_n \leftarrow -\nabla\nabla^T\ln q(\bm{\Gamma})\bigg|_{\bar{\bm{\Gamma}}}$
\EndFor
\end{algorithmic}
\end{algorithm}
\vspace{-10pt}
\section{Simulations and results}

\begin{figure*}[ht]
    \centering
    \includegraphics[width=0.3\linewidth]{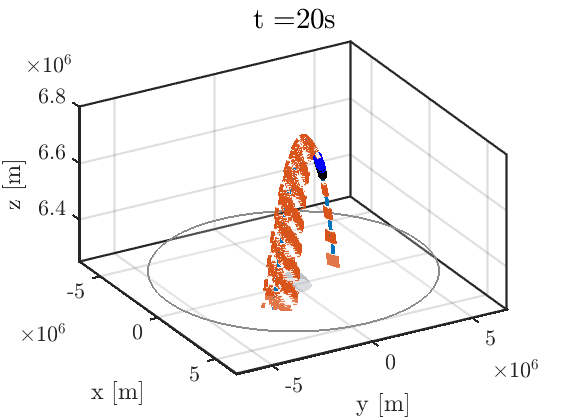}
    \includegraphics[width=0.3\linewidth]{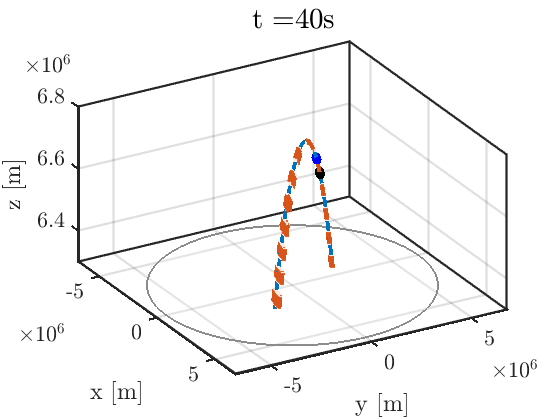}
    \includegraphics[width=0.3\linewidth]{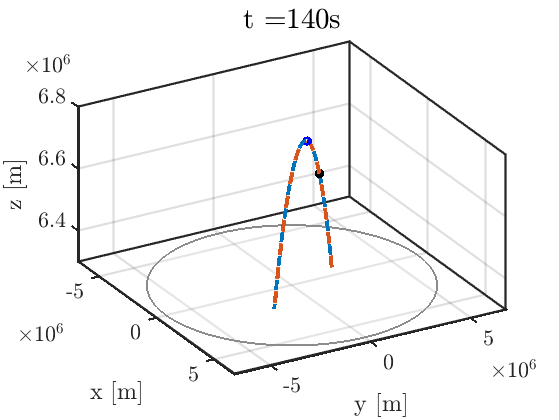}
    \caption{True satellite orbit (blue line) and $95 \% \;CI$ orbit estimation (orange lines) for different time instants.  The starting satellite position is marked with a black dot and the current position is marked with a blue dot.}
    \label{fig:trajectory}
    \vspace{-8pt}
\end{figure*}

\subsection{Simulation setup}
The performance of the proposed method is evaluated in a LEO satellite scenario with the orbit and AoA are generated according to the considered model using \eqref{eq:AoAcalc} with \eqref{eq:prior_Gamma}. Only  parameter settings leading to orbits above the horizon for at least 500\ s are considered. The method is also evaluated on a set of orbits generated by propagating publicly available Starlink TLE data using the Simplified General Perturbations 4 (SGP4) algorithm. As before, only trajectories remaining above the horizon for 500\ s are considered. The signal is generated using \eqref{eq:compact} with analog combining weights chosen as $\bm{b}_n=\bm{a}(\tilde{\bm{\Omega}}_n(\bar{\bm{\Gamma}}))$. The other simulation parameters are presented in Table \ref{tab:parameters}, where Gain indicates the gain of a single antenna element, $P$ is the transmission power of the satellite, $lat/long$ are the coordinates of the GS, and $p$ is the exceedance probability of the atmospheric attenuation. 

The proposed algorithm is compared with a two-step tracking method, where the AoA $\tilde{\bm{\Omega}}(t)$ is tracked using a linear Kalman filter. The AoA is estimated by applying the MUSIC algorithm on a 2.5-degree window around the KF predicted AoA. It was found that the two-step method, needed data every 5 seconds to ensure efficient tracking, whereas the proposed method only requires time updates every 20 seconds.

To evaluate the estimation accuracy, we introduce the mean AoA estimation error as a performance metric defined as
\begin{equation}
   A_e (K) =  \frac{180} {K\pi}\sum_{k=1}^{K}\cos^{-1}\bigl({\tilde{\bm{\Omega}}}^{T}(\bar{\bm{\Gamma}}_n^{(k)})\tilde{\bm{\Omega}}_{n,true}^{(k)}\bigr)
\end{equation}
where $K$ is the number of Monte Carlo runs. For SNR comparisons, the SNR is calculated at $t=0 $.

\begin{table}[t]
    \caption{Simulation parameters}
    \label{tab:parameters}
    \centering
    \begin{tabular}{ccccc}
        \rowcolor{gray!20} 
        \hline
        Carrier freq. & $N_t$ & $M$ & $N_{\text{sub}}$ & Gain \\
        28 GHz & $32\times32$ UPA & $8\times8$ & $4\times 4$ UPA & 5.46 dB \\ \hline
        \rowcolor{gray!20} 
        Antenna type & P & $lat$/ $long$ & $\bm{s}$ & $p$ \\
        isotropic & 5 W & 50.81°/4.38° & Zadoff-Chu seq. & 0.01\% \\ \hline
    \end{tabular}
    \vspace{-15pt}
\end{table}

\begin{figure}[H]
    \centering
    \includegraphics[width=0.95\linewidth]{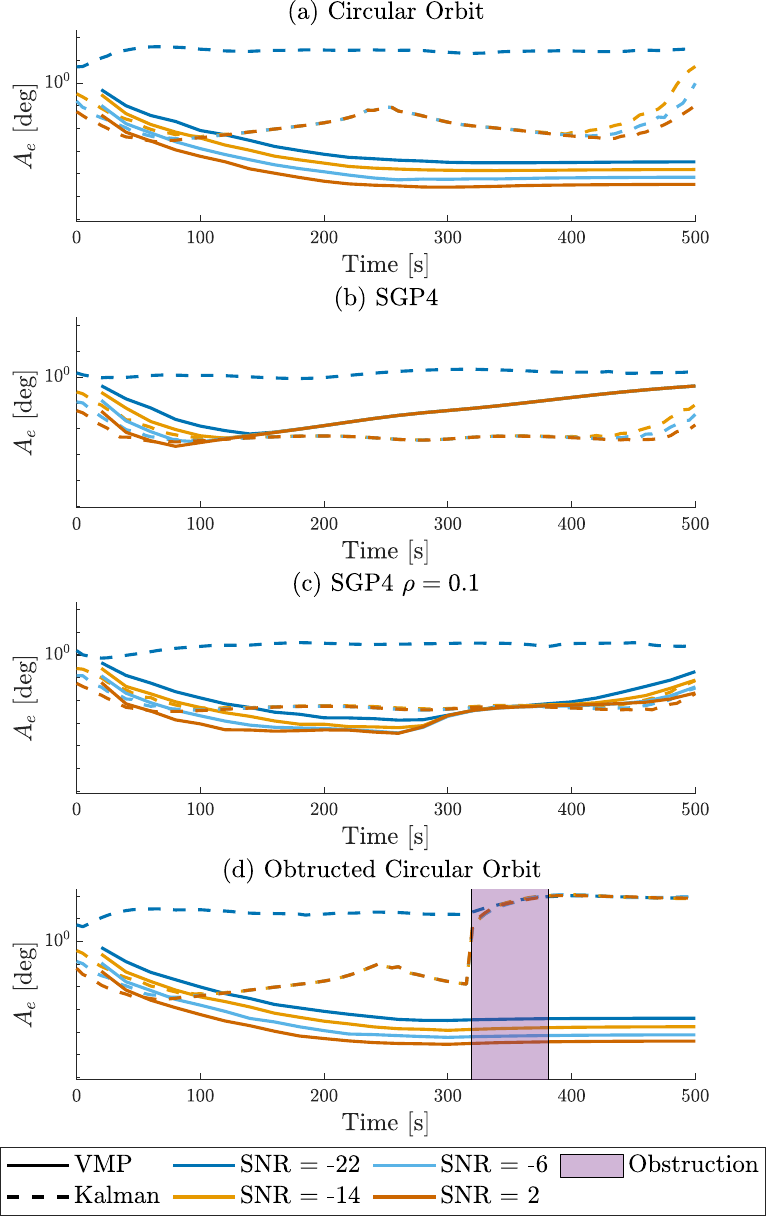}
    \caption{AoA estimation error over time for different SNR values, averaged over $K = 100$ Monte Carlo runs. \textbf{(a)} Unobstructed circular orbits. \textbf{(b)} Unobstructed SGP4  orbits. \textbf{(c)} Unobstructed SGP4 orbits with a time window, $\rho=0.1$ applied to the VMP. \textbf{(d)} Obstructed circular orbits.  }
    \label{fig:monte_carlo_est_error}
\end{figure}

\subsection{Simulation results}

Fig. \ref{fig:trajectory} shows the orbit estimation performance of the VMP algorithm at multiple times, displaying the true orbit together with the orbits corresponding to the 95 $\%$ confidence interval ($95\% \; CI$) of $\bm{\Gamma}$. The $95\% \; CI$ orbits progressively converge toward the true orbit as more data is collected.

Fig. \ref{fig:monte_carlo_est_error} shows the AoA estimation error $A_e$ over time  at different SNRs values, for both circular orbits and orbits generated with SGP4. 

For the circular orbit, the 2-step method presents a higher estimation error than the VMP method. After $400$\ s, the error increases for the 2-step approach as the satellite approaches the horizon due to the lower SNR in this region. The 2-steps approach consistently does not track the satellite at -22dB SNR. In contrast, the estimation error with the VMP decreases over time and achieves a low estimation error across all the tested SNR levels, including -22 dB. 
For the SGP4-generated orbits the VMP is outperformed by the Kalman filter for the entire range given sufficiently high SNR. As all VMP angle errors converge to similar values, independent of SNR, the effect is theorized to be due to model mismatch between the assumed circular orbit and the orbit generated with SGP4. Hence, the data close to the zenith will dominate due to its high SNR and "lock" the estimated orbit. To circumvent this, a window in time is introduced by weighting the sum in \eqref{eq:qGamma} with $\rho^{N-n}$ as seen in Fig.~\ref{fig:monte_carlo_est_error} (c). Here the VMP method again outperforms the two step approach suggesting that for estimating  the SGP4 like orbits, ${\bm{\Gamma}}$, is not fully stationary and to obtain better performance across all orbits, this fact should be expressed in the model.
Fig.~\ref{fig:monte_carlo_est_error} (d) compares $A_e$ for the VMP algorithm and the 2-steps method for a partially-obstructed orbital pass, where we assume that the GS receives only noise, $\mathbf{y}_{n} = \bm{v}_{n}$, from $t = 319\,\mathrm{s}$ to $t = 381 \,\mathrm{s}$. 
During this obstruction period, the AoA error of the 2-steps method increases sharply and fails to recover once the obstruction ends. On the other hand, the AoA error of the proposed algorithm increases slightly, but the algorithm quickly recovers after the obstruction, \color{black}demonstrating its resilience to temporary signal obstruction.

\section{Conclusion}

This paper presents a VMP algorithm for LEO satellite beam tracking, based on the estimation of the parameters of the orbit, considering a hybrid analog-digital beam-forming architecture. Simulation results show that the proposed algorithm successfully estimates these parameters with a few received signals. 
The algorithm allows to estimate the AoA with a low error over a wide range of SNR values, is resilient to obstruction, outperforms classical beam tracking algorithms, and resolves the $2\pi$ AoA ambiguity inherent to hybrid architectures. 
\color{black}

\section*{Acknowledgment}
The work is  funded in part by the COST INTERACT action; in part by the Space4ReLaunch project, which is supported by the SPW Economie Emploi Recherche of the Walloon Region, under the grant agreement nr 2210181; and in part  by the Thomas B. Thriges Foundation grant 7538-1806. 

\bibliographystyle{IEEEtran}
\bibliography{refs}

\end{document}